\documentclass[runningheads]{llncs}
\usepackage[T1]{fontenc}
\usepackage{multirow}
\usepackage{graphicx}
\usepackage{float}
\usepackage{listings}
\usepackage{xcolor}
\usepackage{pifont}
\usepackage{cite}
\usepackage{amsmath,amssymb,amsfonts}
\usepackage{algorithmic}
\usepackage{graphicx}
\usepackage{textcomp}
\usepackage{xcolor}
\usepackage{tabularx}  
\usepackage{listings}
\usepackage{xcolor}
\usepackage{pifont}
\newcommand{\cmark}{\ding{51}} 
\newcommand{\xmark}{\ding{55}} 
\usepackage{listings}

\AtBeginDocument{%
}

\usepackage{listings}
\usepackage{xcolor}

\lstdefinelanguage{Solidity}{
  keywords={
    pragma, solidity, contract, function, modifier,
    mapping, address, public, private, external,
    internal, view, pure, payable, returns,
    return, require, if, else, for, while,
    memory, storage, calldata, event, emit,
    import, struct, enum, bool, string
  },
  sensitive=true,
  comment=[l]{//},
  morecomment=[s]{/*}{*/},
  morestring=[b]",
  morestring=[b]'
}

\begin{document}
%
\title{Solidity Meets LLMs: A Transformer-Based Approach to Smart Contract Vulnerability Detection}

\author{GHORAB Djamel Eddine Hakim\inst{1}\orcidID{0009-0002-5068-0393} \and
MOKHATI Farid\inst{1}\orcidID{0000-0003-4311-342X} \and
GHORAB Mostafa Anouar\inst{2}\orcidID{0000-0003-2235-7126}}

\institute{Research Laboratory on Computer Science’s Complex Systems (RELA(CS)2) 
University of Oum El Bouaghi, Oum El Bouaghi, Algeria
\email{ghorab.djameleddine@univ-oeb.dz}\\ \and
Laval University, Quebec, Canada\\
\email{mostafa-anouar.ghorab.1@ulaval.ca}}

\maketitle 
\markboth{}{}

\begin{abstract}
The growing adoption of blockchain technologies, particularly the Ethereum platform, has amplified the critical role of smart contracts in decentralized applications. However, the increasing complexity and financial value of these contracts make them prime targets for cyber attacks. In this work, we present a transformer-based approach for the detection of vulnerabilities in smart contract fragments written in Solidity. Leveraging the representational power of pre-trained Large Language Models (LLMs), we construct a robust pipeline that includes the definition of a ground truth dataset, labeling code fragments as vulnerable or safe. We then fine-tune a BERT-based architecture on this dataset, enabling the model to capture the syntactic and semantic patterns specific to Solidity code. Our fine-tuned model demonstrates strong performance, achieving an F1 score of 92\%, and highlighting the effectiveness of LLM adaptation in enhancing smart contract security through deep contextual understanding.

\end{abstract}
\section{Introduction}
Blockchain, as a form of distributed ledger technology (DLT) \cite{Punia2024}, provides a secure, transparent,
and immutable mechanism for storing, sharing, and transferring information across a
decentralized network of nodes. It has found extensive usage in critical domains such as finance \cite{Banking9970827},
healthcare \cite{HALEEM2021130}, logistics \cite{10747326}, and notary services \cite{Notarization}, demonstrating impressive results and reducing the
need for third-party intermediaries. 

The advent of smart contracts, which are programs stored and executed on the
blockchain, has further extended the application of blockchain technology to domains such as
the Internet of Things and Edge Computing \cite{Khan2021}. However, like any software, smart contracts are
susceptible to security vulnerabilities. The primary challenge lies in the fact that once it is implemented,
a smart contract is immutable and cannot be modified, making securing a smart contract
during all phases of development a complex task. A notable example of this vulnerability was
seen in the early stages of smart contract deployment when the DAO (Decentralized
Autonomous Organization) was attacked due to a reentrancy vulnerability, leading to the
siphoning off around 70 million US dollars \cite{ZaazaElaBakkali+2023}.

In this paper, we evaluate the power of artificial intelligence and machine learning,
specifically Bidirectional Encoder Representations from Transformers (BERT) \cite{devlin2019bertpretrainingdeepbidirectional} to address the critical issue of security vulnerabilities in smart contracts written in Solidity.
Our work involves a comprehensive process that begins with the collection of raw
solidity smart contracts. These contracts are then labeled, forming a dataset that serves as the
foundation for our model training. Once this preparatory phase is complete, we proceed to train
our BERT model, fine-tuning it to accurately detect potential vulnerabilities in fragments of smart
contracts.

This paper is organized as follows: we start by providing background information on Solidity smart contracts. Next, we discuss the different vulnerabilities found in Solidity contracts and introduce a simplified taxonomy. We then review previous approaches aimed at detecting and mitigating these vulnerabilities, ranging from formal methods and software engineering techniques to machine learning-based solutions. Finally, we present our methodology for building a model capable of detecting vulnerabilities in Solidity smart contracts code fragments.

The remainder of this paper is organized as follows. Section 2 provides background information on Solidity smart contracts. In section 3, we give a brief overview of major related works to our approach.  Section 4 presents the approach we proposed for building a model capable of detecting vulnerabilities in Solidity smart contracts  code fragments.
Finally, we give a conclusion and some future work directions in Section 5.

\section{Background}
\subsection{Solidity Smart Contract}
Solidity Smart Contracts are program stored on the Ethereum Blockchain, and their execution is enforced through the consensus protocol. It represents a set of rules written in a programming language. That indicates specified actions will be executed when certain conditions are met. As an Object-Oriented Programming Language, Solidity supports inheritance, overloading, and overriding. It also has unique features such as modifiers, specifiers and global variables related to the Ethereum Blckchain and the Ethereum Virtual Machine (EVM).

\begin{table}[h]
    \centering
    \caption{Example of Ethereum variables}
    \begin{tabularx}{\linewidth}{|l|l|X|}
        \hline
        \textbf{Variable} & \textbf{Type} & \textbf{Description} \\ \hline
        \texttt{block.number} & \texttt{uint} & Current block number \\ \hline
        \texttt{block.timestamp} & \texttt{uint} & Current block timestamp as seconds since the Unix epoch \\ \hline
        \texttt{msg.sender} & \texttt{address} & Address of the message sender \\ \hline
        \texttt{msg.value} & \texttt{uint} & Number of wei sent with the message \\ \hline
        \texttt{tx.origin} & \texttt{address} & Sender of the transaction (full call chain) \\ \hline
    \end{tabularx}
    \label{tab:ethereum_variables}
\end{table}

\subsection{Vulnerabilities and Taxonomy}
As the landscape of blockchain technology and smart contracts evolves, so does the complexity and variety of potential security vulnerabilities. And with extra features come extra problems. Solidity contracts, due to their critical use in finance, health, and other sensitive applications, are susceptible to traditional vulnerabilities, as well as new attack vectors inherent to the nature of smart contracts.
To assist developers, auditors, and researchers in navigating this challenging terrain, we present, as shown in Table~\ref{tab:vulnerabilities}—a taxonomy of Solidity smart contract vulnerabilities. This classification is designed to encapsulate the broad spectrum of issues identified over time, distilled into five main categories that reflect their impact and behavior. 

\begin{table}[h]
    \centering
    \caption{Taxonomy of vulnerabilities in Solidity Smart Contracts}
    \begin{tabular}{|l|p{7cm}|}
        \hline
        \textbf{Category} & \textbf{Impact} \\
        \hline
        Cross-Call Vulnerabilities & These vulnerabilities can result in unauthorized actions or unintended modifications to a contract's state because of unforeseen external calls. A classic illustration of this is the reentrancy vulnerability. \\
        \hline
        Denial of Service (DoS) Vulnerabilities & These vulnerabilities affect the availability of a smart contract, hindering it from performing its intended operations. Often results from conditions that consume all available gas or use unlimited iterations, rendering the contract inoperable. \\
        \hline
        Arithmetic Vulnerabilities & Improper handling of arithmetic operations can lead to financial loss or corrupted contract logic. This category includes issues like integer overflow and underflow, division by zero, and rounding errors. \\
        \hline
        Access Control Vulnerabilities & These vulnerabilities can allow unauthorized users to access restricted functions or data within a contract. A typical example would be the use of tx.origin for authorization. \\
        \hline
        Bad Randomness Vulnerabilities & Reliance on weak or predictable randomness sources can compromise the fairness and security of a contract’s operations. \\
        \hline
    \end{tabular}
    \label{tab:vulnerabilities}
\end{table}

The Reentrancy vulnerability, a typical example of a cross-call vulnerability, gained notoriety during the infamous Decentralized Autonomous Organization (DAO) exploit in 2016, where malicious actors siphoned 3.6 million ETH, worth approximately 70 million USD at the time, from the DAO smart contract.

\begin{lstlisting}[language=solidity, caption=SimpleDAO with Reentrancy vulnerability,label={lst:SimpleDAO}, captionpos=b, numbers=none,float]
contract SimpleDAO {
    mapping(address => uint) public credit;
    bool public flag; bytes data;
    function donate(address to) public payable {
        credit[to] += msg.value;
    }
    function withdraw(uint amount) public {
        if (credit[msg.sender] >= amount) {
            (flag, data) = msg.sender.call.value(amount)("");
            if (flag == true) {
                credit[msg.sender] -= amount;
            }
        }
    }
}
\end{lstlisting}

Looking at the withdrawal function in the listing \ref{lst:SimpleDAO}, if the credit balance of the sender is greater than
or equal to the specified withdrawal amount, then the contract will initiate an external call to
the \textit{msg.sender} to transfer the specified amount of ether. When the external call is executed,
the control will be transferred to the recipient and they can execute arbitrary code. Listing \ref{lst:SimpleDAOExploit} represents a malicious smart contract designed to exploit the Reentrancy vulnerability in listing \ref{lst:SimpleDAO}, the transfer of ether will invoke the attacker's receive function, which
will recall the withdraw function of the vulnerable smart contract. In this way, the attacker will be
capable of siphoning Ether and emptying the balance of the vulnerable contract.

\begin{lstlisting}[caption=Reentrancy Attack Contract written in Solidity,label={lst:SimpleDAOExploit},captionpos=b , numbers=none,float]
interface SimpleDAO {
    function donate(address to) external payable;
    function withdraw(uint amount) external;
}
contract ReentrancyAttacker {
    SimpleDAO public vulnerableDAO;
    constructor(address _vulnerableDAO) public {
        vulnerableDAO = SimpleDAO(_vulnerableDAO);
    }
    function attack() public payable {
      vulnerableDAO.donate{value: msg.value} address(this));
      vulnerableDAO.withdraw(msg.value);
    }
    receive() external payable {
        if (address(vulnerableDAO).balance >= 0.01 ether) {
            vulnerableDAO.withdraw(0.01 ether);
        }
    }
}
\end{lstlisting}

Remediation can be performed using the \textit{transfer} method instead of making an external call.
Alternatively, using a lock to ensure that the withdrawal function will not be called until the
balance is updated.

\section{Related works}
The security of blockchain technologies-particularly smart contracts written in solidity-has attracted considerable attention over the past decade. Multiple approaches have been proposed to analyze, verify and enhance the security of smart contracts prior to deployment \cite{Vidal_2024}. 

\subsection{Detection approaches}
Nehaï et al. proposed a modeling method for smart contracts written for the Ethereum
Blockchain \cite{8726806}. They started by modeling the Ethereum Blockchain as a distributed system for
managing transactions between different clients. Next was proposing different rules for
translating a smart contract written in Solidity into the NuSMV input language. Properties to
be verified are formalized into computation tree logic (CTL). Moreover, testing whether a certain type
of vulnerability exists requires model checking of the correspondent properties. 

Building on dynamic techniques for vulnerability detection, Jiang et al. developed a fuzzing tool that analyzes the application binary interface(ABI) and bytecode of a smart contract. The tool extracts data types and function signatures from the ABI, then generates fuzzing inputs accordingly. During the fuzzing process, the tool monitors logs to identify security vulnerabilities triggered by the inputs \cite{9000089}.

In contrast to fuzzing, which uses concrete input values, symbolic execution explores program paths using symbolic values. It employs a Symbolic Execution Engine that traverses control flow paths (CFPs) by interpreting symbolic inputs until it reaches expressions formulated in terms of those symbols. A model checker or solver is then used to determine whether any security properties are violated along the explored paths \cite{10.1145/3182657}.

An example of symbolic execution in practice is the Osiris framework that uses symbolic execution to find integer bugs in Solidity Smart contracts\cite{10.1145/3274694.3274737}. Osiris starts by inferring type information about integers
such as size and signedness (e.g., uint32 has a 32-bit size and is unsigned) from the bytecode. After that, for each instruction that could potentially overflow (i.e., ADD or MUL) or underflow (i.e., SUB), a constraint will be constructed, and a solver will check whether that constraint is satisfied or not. The tool will signal a bug as valid if it originates from a source and flows towards a sink.

Beyond dynamic and symbolic methods, static analysis techniques such as abstract interpretation, pattern matching, and taint analysis are also widely used for smart contract vulnerability detection \cite{10.1145/3274694.3274737,8445052}.

\begin{table}[htbp]
\centering
\caption{Summary of Detection Tools and Techniques}
\label{tab:categories_techniques_tools}
\begin{tabular}{|l|l|l|}
\hline
\textbf{Category} & \textbf{Technique} & \textbf{Tool} \\
\hline

Formal Verification
  & Model Checking
  & \begin{tabular}[c]{@{}l@{}}
      FSolidM \cite{mavridou2018tooldemonstrationfsolidmdesigning}\\
      EthSemantics\\
      SmartPulse\\
      VeriSmart\\
      VeriSolid
    \end{tabular} \\
\hline

\multirow{2}{*}{Software Testing}
  & Fuzzing
  & \begin{tabular}[c]{@{}l@{}}
      ContractFuzzer\\
      OYENTE\\
      Echidna\\
      sFuzz\\
      EthRacer\\
      Solanalyser\\
      sFuzz\\
      GasFuzzer\\
      xFuzz\\
      Etherolic
    \end{tabular} \\
\cline{2-3}

  & Symbolic Execution
  & \begin{tabular}[c]{@{}l@{}}
      Osiris\\
      Oyente\\
      DEPOSafe\\
      SAILFISH\\
      sCompile\\
      SmartScopy\\
      teEther\\
      Vultron
    \end{tabular} \\
\hline

\multirow{3}{*}{Static Analysis}
  & Abstract Interpretation
  & \begin{tabular}[c]{@{}l@{}}
      Securify\\
      SoliDetector\\
      Vandal\\
      Zeus
    \end{tabular} \\
\cline{2-3}

  & Pattern Recognition
  & \begin{tabular}[c]{@{}l@{}}
      SmartCheck\\
      NeuCheck\\
      SolidityCheck\\
      Vrust
    \end{tabular} \\
\cline{2-3}

  & Taint Analysis
  & \begin{tabular}[c]{@{}l@{}}
      Clairyoyance\\
      EasyFlow\\
      Ethainter\\
      EthPloit\\
      Osiris\\
      Sereum\\
      Slither
    \end{tabular} \\
\hline

Artificial Intelligence
  & Machine and Deep Learning
  & \begin{tabular}[c]{@{}l@{}}
      SoliAudit\\
      SmartMixModel
    \end{tabular} \\
\hline

\end{tabular}
\end{table}

In recent years, machine learning and deep learning techniques have been increasingly applied to both source and bytecode representations of smart contracts. For example, SmartMixModel uses features extracted from high-level source code and low-level bytecode to perform binary classification of multiple types of vulnerabilities\cite{9881798}. Similarly, Zhuang et al. utilized the Abstract Syntax Tree (AST), Call Graph(CG), and Call Flow Graph(CFG) of a smart contract to construct a representative graph. This graph was subsequently segmented based on criteria related to the type of vulnerability. For example, in the case of integer overflow vulnerabilities, the graph was sliced according to arithmetic operations. Following this, the sliced graph is then fed into a Graph Neural Network (GNN) for feature extraction and vulnerability classification. \cite{ijcai2020p454}.

Another deep learning-based approach involves the use of Convolutional Neural Networks (CNNs). Here, the bytecode of a smart contract is transformed into an RGB image, allowing CNNs to be used for classification of vulnerabilities based on patterns in the resulting images \cite{BlockSys2022}.

\subsection{Comparison of Solidity Smart Contract Vulnerability Detection Tools}

In an exploration of several Solidity scanners, this assessment evaluates their
effectiveness based on key criteria crucial during the development phase of smart contracts.
The focus is on understanding the tools' requirements for binary code, reliance on Abstract
Syntax Trees (AST), and their ability to function with partial code during the development
phase.

Most tools heavily rely on either the Binary Code and Application Binary Interface (ABI) or
the Abstract Syntax Tree (AST). Tools like SmartCheck have their own proper parsers,
requiring a complete smart contract to be parsed into an Intermediary Representation (IR). This
characteristic makes them unsuitable for scanning parts of the code and integrating them into
integrated development environments (IDE).

On the other hand, tools such as SolScan utilize regular expressions to identify vulnerable patterns. Although
this approach allows SolScan to detect vulnerabilities in partial code, it comes with limitations.
Regular expression rules are rigid and may not effectively uncover newer patterns of vulnerabilities. This may result in SolScan missing out on detecting emerging vulnerabilities in
smart contracts. Table \ref{tab:code_analysis_tools} demonstrates the detection level of the tool, whether it works on source code or byte code, and also the requirements for either the bytecode, the abstract syntax tree, or an unfinished source code.

\begin{table*}[h]
    \centering
    \caption{Comparison of Code Analysis Tools}
    \begin{tabular}{lccccc}
        \hline
        \textbf{Tool} & \textbf{Detecting Level} & \textbf{Bytecode} & \textbf{AST} & \textbf{Type} & \textbf{Unfinished Code} \\
        \hline
        SmartPulse & Source Code & \xmark\ & \xmark\ & Static & \xmark\ \\
        ContractFuzzer & Bytecode &  \cmark\ & \xmark\ & Dynamic & \xmark\ \\
        SolScan & Source Code & \xmark\ & \xmark\ & Static & \cmark\ \\
        Osiris & Bytecode & \cmark\ & \xmark\ & Static & \xmark\ \\
        SmartCheck & Source Code & \xmark\ & \xmark\ & Static & \xmark\ \\
        Slither & Source Code & \xmark\ & \cmark\ & Static & \xmark\ \\
        \hline
    \end{tabular}
    
    \label{tab:code_analysis_tools}
\end{table*}

Unlike traditional methods such as model checking, fuzzing, symbolic execution, and rule-based static analysis - which depend heavily on hand-crafted constraints, formal rules, or symbolic input - our approach provides a more adaptable and generalizable solution. Using the bidirectional contextual understanding of the BERT model, it can effectively analyze code snippets without the need for explicit formal modeling or manual rule definition. This makes it especially well-suited for detecting a broad range of vulnerability patterns across various stages of development, including early-stage or incomplete code, where conventional tools often fall short due to their reliance on abstract syntax tree, fully compiled code or well-specified inputs.
\section{The proposed approach}
The integration of large language models (LLM) techniques into software engineering tasks has yielded
remarkable results, spanning from code generation to test case creation and code analysis. In
our work, we aim to leverage the power of an NLP model, BERT, which has demonstrated
exceptional performance across various software engineering applications \cite{chaieb2024detectingandroidmalwareneural}. Specifically, we
intend to fine-tune BERT for the binary classification of Solidity source code, determining
whether it contains vulnerabilities or not. This approach capitalizes on BERT's capabilities to
understand the intricacies of programming languages and identify potential security
vulnerabilities.

\begin{figure}[htbp]
    \centering
    \includegraphics[width=0.9\linewidth]{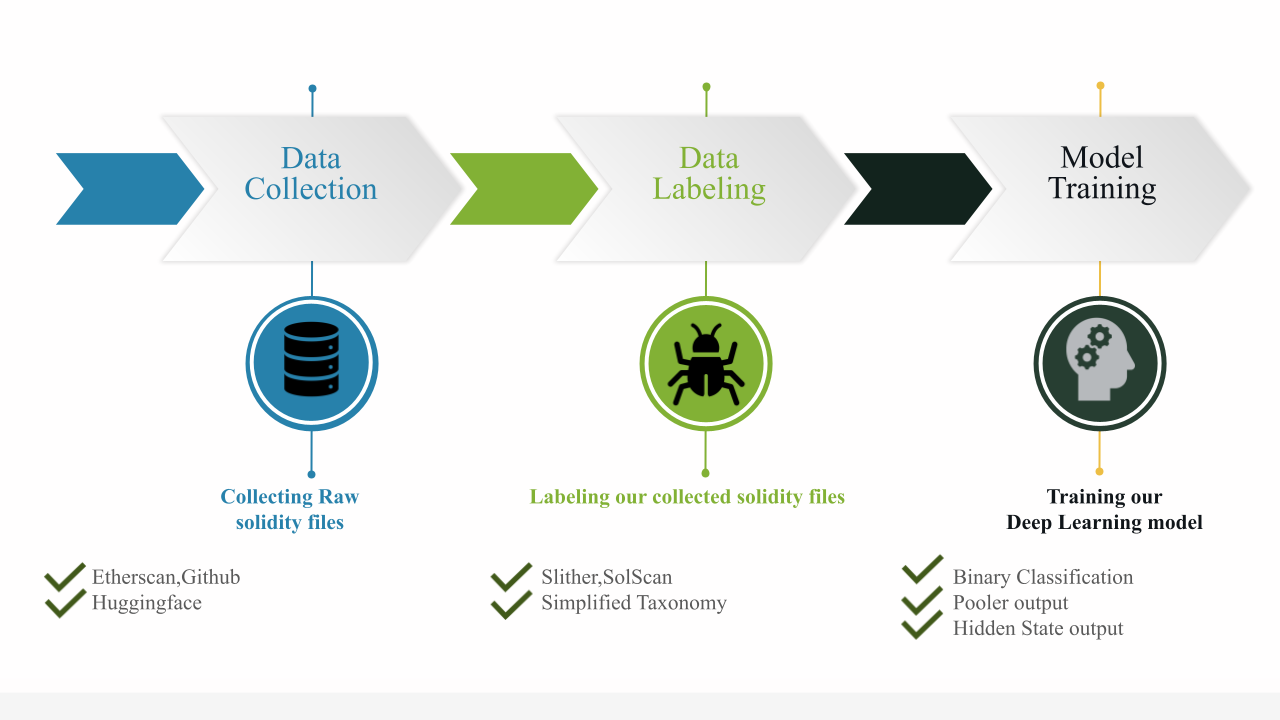}
    \caption{Global view of our approach}
    \label{fig:methodology}
\end{figure}

Our approach spans from data collection to automating the labeling process, to fine-tuning the BERT model for binary classification of Solidity code fragments, as illustrated in Figure~\ref{fig:methodology}.
\vspace{1,5em}
\vspace{-2.5em}
\subsection{Data collection and Labeling process} 
\vspace{-0.5em}
In our data collection phase, we began by gathering raw Solidity smart contract files from
various resources, such as previous works \cite{rossini2022slitherauditedcontracts,10.1145/3543507.3583367} and contracts verified on Etherscan. These
smart contracts were either unlabeled or labeled heterogeneously, depending on the taxonomy
used by different researchers. Due to the labor-intensive and time-consuming nature of manual
labeling, we opted to use existing tools, namely Slither \cite{Feist_2019} and SolScan, to classify code fragments as our ground truth. Due to the criticality of smart contracts and following the principle of "better safe than sorry," a detection by either tool is considered indicative of vulnerabilities, whereas a smart contract is deemed safe only if both tools classify it as such.
\vspace{1em}

\begin{figure}[H]
    \centering
    \includegraphics[width=0.9\linewidth]{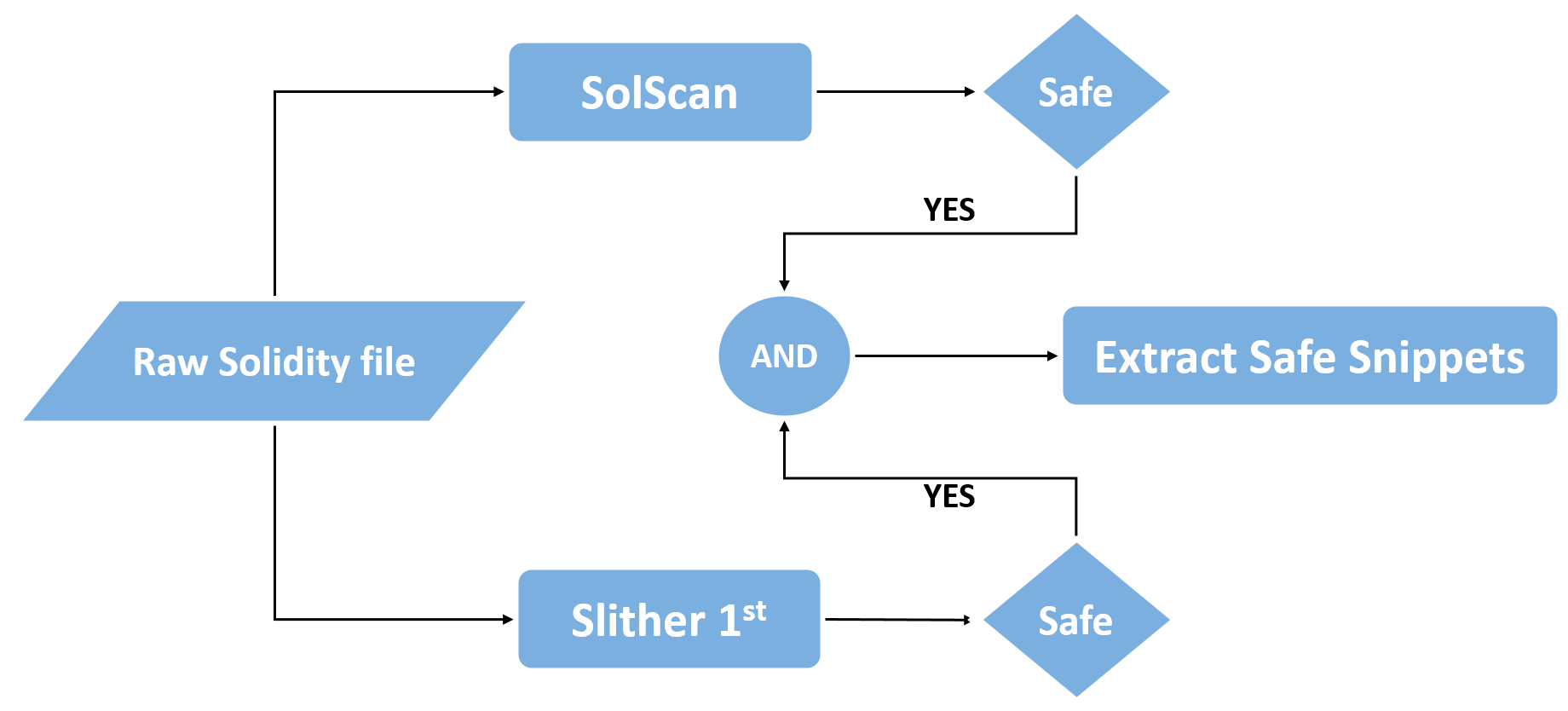}  
    \caption{Labeling Safe Code Fragments}
    \label{fig:labelingsafe}
\end{figure}

Slither and SolScan employ different approaches for vulnerability detection. Slither uses symbolic execution, while SolScan utilizes pattern recognition. For each smart contract, we used both tools to scan and determine the safety of the smart contract. A contract is deemed safe only if both tools classify it as such. In such cases, we extract snippets from the smart contract and label them as safe. The process is illustrated in Fig.~\ref{fig:labelingsafe}.

\begin{figure}[H]
    \centering
    \includegraphics[width=0.9\linewidth]{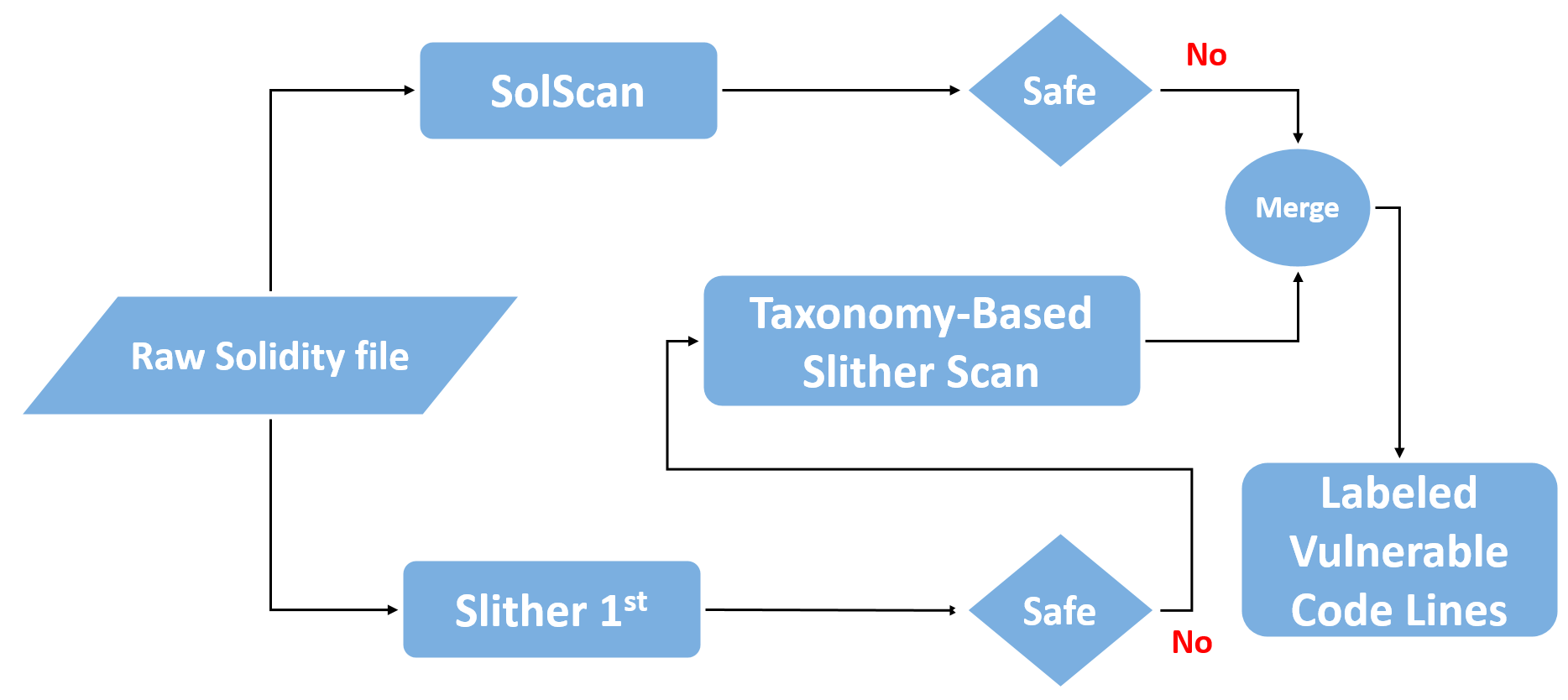}  
    \caption{Labeling Vulnerable Code Fragments}
    \label{fig:labelingvul}
\end{figure}
\vspace{-1em}

On the other hand, if SolScan or Slither detects the smart contract as vulnerable, we extract the vulnerable code fragment based on the results of the scanners. 
Figure \ref{fig:labelingvul} demonstrates the process for extracting vulnerable code fragments.

\begin{figure}[!t]
    \centering
    \includegraphics[width=0.9\linewidth]{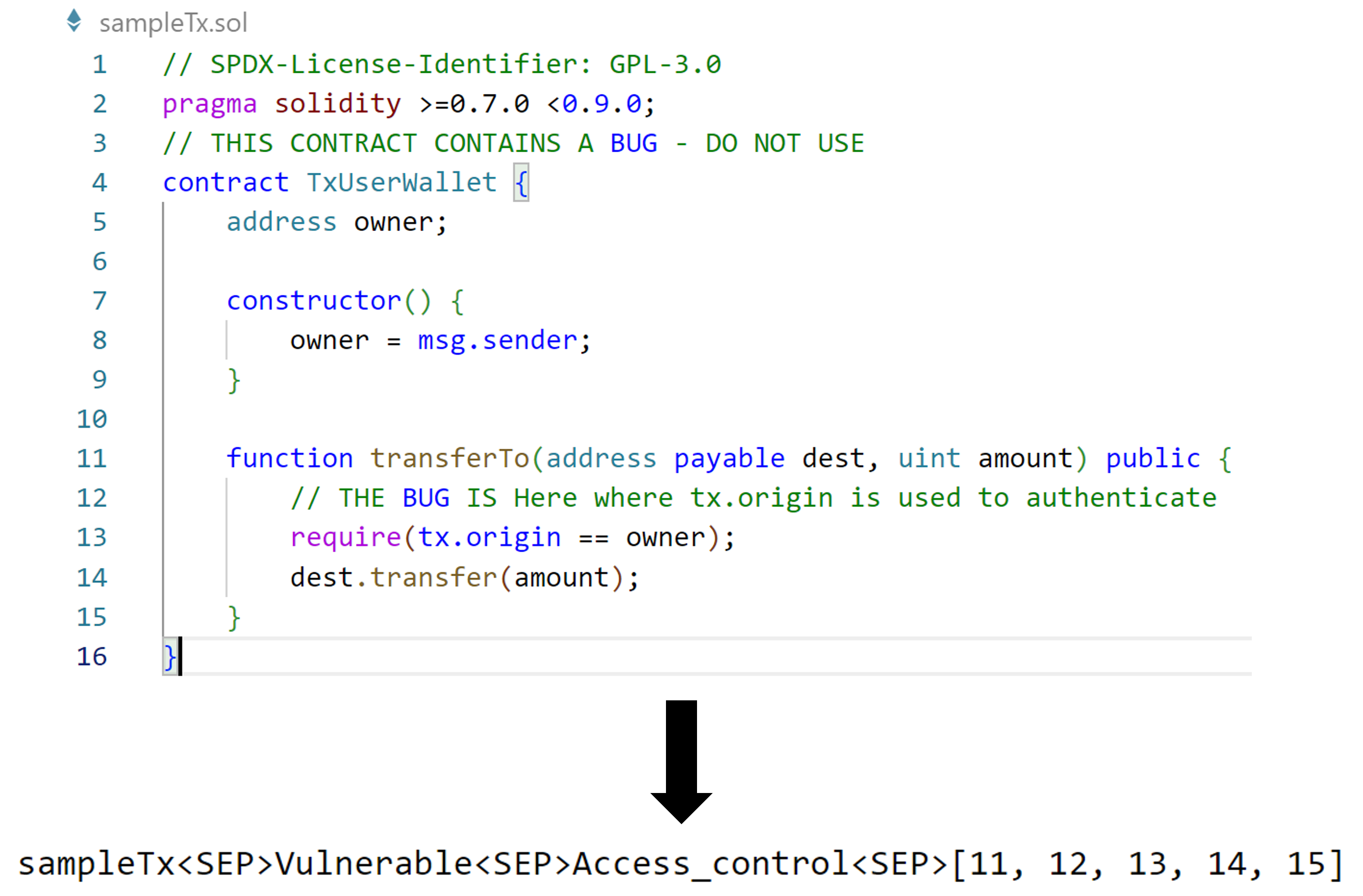}  
    \caption{Extraction of Vulnerable lines from a Solidity Smart Contract}
    \label{fig:Extract}
\end{figure}

In the first slither scan, the smart contract is analyzed by all Slither's modules, We then perform a second scan using multiple specific Slither modules to detect the type of vulnerability according to our taxonomy. This taxonomy categorizes Solidity vulnerabilities into arithmetic, access control, cross-call vulnerabilities, and bad randomness vulnerabilities. As demonstrated in Fig. ~\ref{fig:Extract}, we analyze Slither’s output and save the line numbers inferred as the source of vulnerability. Our modified version of SolScan, based on our taxonomy, automatically saves
the detected line numbers along with the vulnerability type without needing another scan. Upon
completing the scans using Slither and SolScan, we combine the results. Our data is then
structured as : 
\begin{center}
\texttt {contract\_name<SEP>Safety<SEP>Type<SEP>Lines} \end{center}
 as demonstrated in Fig. \ref{fig:Extract}.

Finally, we extract the actual code. To do this, we read each smart contract and extract
the vulnerable lines. After extraction, we treat the code snippet by removing simple comments,
multiple-line comments, extra spaces, and new lines. 

\subsection{Training}
Before initiating the training process, we started a preparatory phase that involved
loading our data and partitioning it into distinct subsets: training, validation and testing data.

The training set, which comprises 70\% of the total data, served as the primary source for the model to learn and adapt. It was through exposure to this dataset that the model was able to adjust its internal parameters and develop its predictive capabilities.
The validation set, which makes up 15\% of the data, used to follow our model and mitigate the risk of overfitting. By evaluating the performance of the model on the validation set during the training phase.
The remaining 15\% data were reserved as the testing set. This data set was used exclusively to assess the performance of the model in unseen data, providing an evaluation of the final model.

Once our data was properly partitioned, we proceeded to the preprocessing stage. This
involved transforming our raw data into a format that could be effectively ingested by the
BERT model. To accomplish this, we used the BertTokenizer from the Hugging Face
Transformers library. This tokenizer converted our text data into tokenized form, suitable as an
input into the BERT model.

For the actual training process, we used the computational power of an \textbf{A100-
SXM4-40GB} GPU with \textbf{50GB} of RAM.

\subsection{Results}

In this section, we present the results of our model. The model was trained over five epochs. As can be seen in Table \ref{tab:metrics}, the performance of the model improved with each epoch. Training and validation losses decreased, indicating that the model was effectively learning from training data and generalizing well to unseen data. The accuracy, precision, recall, \textbf{f1-score}, \textbf{mcc} (Matthews correlation coefficient), and \textbf{roc\_auc} (Area Under the Receiver Operating Characteristic Curve) all increased, demonstrating the increasing proficiency of the model in correctly classifying data.

\begin{table*}[h]
    \centering
    \caption{Performance Metrics Across Five Training Epochs}
    \begin{tabular*}{\textwidth}{@{\extracolsep{\fill}}ccccccccc}
        \hline
        Epoch & Training Loss & Validation Loss & Accuracy & Precision & Recall & F1-score & MCC & Roc\_AUC \\
        \hline
        1 & 0.551955 & 0.382489 & 0.92375 & 0.931871 & 0.92375 & 0.923976 & 0.847273 & 0.924189 \\
        2 & 0.412967 & 0.283605 & 0.93500 & 0.945209 & 0.93500 & 0.935854 & 0.869422 & 0.937985 \\
        3 & 0.361043 & 0.244674 & 0.94000 & 0.945190 & 0.94000 & 0.939730 & 0.875449 & 0.936440 \\
        4 & 0.339168 & 0.228827 & 0.94375 & 0.951237 & 0.94375 & 0.943672 & 0.886339 & 0.942216 \\
        5 & 0.323758 & 0.224617 & 0.94375 & 0.949393 & 0.94375 & 0.943279 & 0.887873 & 0.940091 \\
        \hline
    \end{tabular*}
    \label{tab:metrics}
\end{table*}

These results provide strong evidence of our model’s effectiveness in detecting vulnerable code
in Solidity. The model’s high precision and recall scores indicate its capability to accurately
identify vulnerabilities with a low rate of false positives and false negatives. The high F1-score
and MCC further confirm the model’s robustness.
Our model has demonstrated promising results in the task of detecting vulnerable code in
Solidity, making it a valuable approach for developers and security analysts in the field of smart
contract development. We believe that with further tuning and optimization, the performance
of the model can be significantly improved.

\section{Conclusion and perspectives}
Blockchain technology has significantly transformed various sectors by providing a secure, decentralized, and immutable framework for data and transaction management. The advent of smart contracts has further enhanced these capabilities by enabling automated, trustless transactions without the need for third-party intermediaries. However, the development
of smart contracts is not without its challenges. One of the primary concerns is the potential security vulnerabilities that could compromise smart contracts and, by extension, the systems that utilize them. Therefore, the detection and mitigation of these vulnerabilities in smart contracts have become a critical procedure.

In this study, we proposed a vulnerability detection approach based on Large Language Models (LLMs), specifically BERT, to classify Solidity smart contract code snippets as either vulnerable or safe. Our approach involved collecting and labeling a dataset of Solidity smart contract fragments, training a transformer-based model, and evaluating its ability to detect potential security flaws.

Although our findings validate the effectiveness of LLMs in smart contract vulnerability detection, several directions can be pursued to further improve performance and applicability:

\begin{itemize}
\item \textbf{Enhanced Dataset Quality}: Incorporating corrected vulnerable examples, misleading code, and edge cases will enrich the dataset and reduce false positives.
\item \textbf{Improved Ground Truth Labeling}:Combining static and dynamic analysis tools, complemented by expert manual review, can yield more precise and trustworthy labels.
\item \textbf{Fine-Grained Vulnerability Detection}: Moving from binary to multi-class classification using a comprehensive vulnerability taxonomy can offer deeper insights and assist developers in understanding specific flaws.
\item \textbf{Toward Automatic Remediation}: Utilizing advanced LLMs not only for detection but also for generating secure fixes opens the door to automated vulnerability mitigation.
\end{itemize}

These enhancements will contribute to building a more robust pipeline for secure smart contract development, bridging the gap between vulnerability detection and automated remediation.

\begin{credits}
\subsubsection{\ackname} This research work was supported by the General Direction of Scientific Research and Technological Development (DGRSDT) of the Ministry of Higher Education and Scientific Research of Algeria. The authors thank DGRSDT for the achievement of this work.

\subsubsection{\discintname}
The authors have no competing interests to declare that are relevant to the content of this article.
\end{credits}
%
%
%
\bibliographystyle{splncs04}
%

\bibliography{references}

\end{document}